\PassOptionsToPackage{dvipsnames}{xcolor}
\documentclass[aip,aplmat,reprint,a4paper,superscriptaddress,floatfix,amsmath,amssymb,amsfonts,noshowpacs,longbibliography]{revtex4-2}

\usepackage{newtxtext,newtxmath}
\usepackage[T1]{fontenc}
\usepackage[utf8]{inputenx}
\input{ix-utf8enc.dfu}
\usepackage{graphicx}
\usepackage{sidecap}
\sidecaptionvpos{figure}{t}
\usepackage[dvipsnames]{xcolor}
\usepackage{soul} 
\usepackage{xspace}
\usepackage{siunitx}
\usepackage{url}

\usepackage{floatrow}

\usepackage[
,textwidth=17.5cm
,textheight=23.7cm
,verbose
,pdftex
]{geometry}
\usepackage{xspace}

\usepackage[pdftex]{hyperref}
\hypersetup{
	pdftitle={Crack-free Sc$_x$Al$_{1-x}$N(000\=1) layers grown on Si(111) by plasma-assisted molecular beam epitaxy},
	colorlinks,
	citecolor=blue,
	linkcolor=blue,
	urlcolor=blue
}

\newcommand{\ScAlN}{Sc$_x$Al$_{1-x}$N\xspace}
\newcommand{\ten}{Sc$_{0.1}$Al$_{0.9}$N\xspace}
\newcommand{\twenty}{Sc$_{0.2}$Al$_{0.8}$N\xspace}
\newcommand{\twentyfive}{Sc$_{0.25}$Al$_{0.75}$N\xspace}
\newcommand{\thirty}{Sc$_{0.3}$Al$_{0.7}$N\xspace}
\newcommand{\temp}{\,°C\xspace}
\newcommand{\tw}{2$\uptheta-\upomega$\xspace}
\DeclareSIUnit\angstrom{\text {Å}}
\newcommand{\Tg}{$T_\text{G}$\xspace}

\begin{document}

\title{Crack-free Sc$_{\boldsymbol{\mathsf{x}}}$Al$_{\boldsymbol{\mathsf{1-x}}}$N(000\=1) layers grown on Si(111) by plasma-assisted molecular beam epitaxy}

\author{Duc V. Dinh}
\email[Electronic email: ]{duc.vandinh@pdi-berlin.de}
\affiliation{Paul-Drude-Institut für Festkörperelektronik, Leibniz-Institut im Forschungsverbund Berlin e.V.,\\ Hausvogteiplatz 5--7, 10117 Berlin, Germany.}

\author{Zhuohui Chen}
\thanks{$\dagger$ Deceased on November 4, 2024}
\affiliation{Huawei Technologies Canada Co., Ltd., 303 Terry Fox Drive, Kanata, Ontario, K2K 3J1, Canada.}

\author{Oliver Brandt}
\affiliation{Paul-Drude-Institut für Festkörperelektronik, Leibniz-Institut im Forschungsverbund Berlin e.V.,\\ Hausvogteiplatz 5--7, 10117 Berlin, Germany.}


\begin{abstract}
We investigate the synthesis of 340-nm-thick \ScAlN layers with $0 \leq x \leq 0.35$ on AlN-buffered Si(111) by plasma-assisted molecular beam epitaxy. We employ an AlN nucleation layer under conditions giving rise to single-domain N-polar [(000\=1)-oriented] layers, as demonstrated by the ($3 \times 3$) pattern observed in reflection high-energy electron diffraction and confirmed by KOH etching. The subsequent growth of pure wurtzite Sc$_x$Al$_{1-x}$N layers with $x \leq 0.1$ is feasible at temperatures $\leq$\,740\temp. However, layers with $x \geq 0.2$ grown at 740\temp develop cracks due the high thermal mismatch between \ScAlN and Si. Lowering the growth temperature to 500\temp not only prevents cracking but also improves the crystallinity of the layers. For \thirty layers grown at 500\temp, additional x-ray reflections due to intermetallic AlSc and Al$_3$Sc inclusions are observed. The formation of these compounds can be inhibited by lowering the temperature further to 300\temp.
\end{abstract}

\maketitle

Wurtzite \ScAlN alloys have recently attracted much interest because, as long as the wurtzite structure of \ScAlN is maintained, their longitudinal piezoelectric coefficient $d_{33}$ along the [0001] axis has been reported to strongly increase with increasing Sc content $x$.\cite{Akiyama2009Feb} \ScAlN has thus been used primarily for applications in surface-acoustic-wave (SAW) devices\cite{Hashimoto2013Mar,Campos2018,Park2020,Qamar2020,Yuan_2024} and field-effect transistors,\cite{Hardy2017,Ligl2020May,Wang2021Aug,Krause2022Nov,Dinh2023} but also for ferroelectric devices \cite{Fichtner2019,Wang2022,DWang2022,Wang2023} and distributed Bragg reflectors.\cite{vanDeurzen2023Dec} The integration of \ScAlN with Si opens up opportunities to improve the functionalities of Si-based devices. In fact, \ScAlN/Si has been utilized in diverse applications including SAW devices,\cite{Campos2018,Park2020,Qamar2020} the seamless integration of piezoelectric and piezoresistive devices,\cite{Qamar2020} as well as photonic circuits.\cite{Zhu2020}

Epitaxial growth of uniform and phase-pure \ScAlN layers on Si(111) is challenging because of several reasons. First of all, \ScAlN and Si(111) exhibit a high lattice mismatch (19\% for $x=0$) giving rise to a higher density of threading dislocations than on substrates such as Al$_2$O$_3$(0001) or SiC(0001). Furthermore, the reactivity of Si with the activated N$^*$ from the plasma cell requires specific strategies to avoid the formation of amorphous Si$_x$N$_y$ during the nucleation stage. This stage is also critical in determining the polarity of the \ScAlN layer, since Si in principle allows the formation of both Al and N polar domains.\cite{Lebedev1999Dec,Dasgupta2009,Ledyaev} Second, \ScAlN and Si(111) are also highly thermally mismatched (about 70--75\% for $x=0$),\cite{Lu2018} inducing tensile stress during cooling from growth to room temperature, which is likely to generate cracks. Third, the binary constituents of \ScAlN crystallize in different equilibrium structures, namely, wurtzite for AlN and rocksalt for ScN. \ScAlN is thus prone to phase separation for higher Sc contents, eventually destabilizing the wurtzite structure altogether.\cite{Hoglund2009Jun,Hardy2020,Dinh2023} In addition, intermetallic compounds such as Al$_3$Sc (Cu$_3$Au structure) and AlSc (CsCl structure) may form particularly for metal-stable conditions and high growth temperatures.\cite{Elliott1981Sep,Schuster1985Jul} For the above mentioned applications, it is imperative to maintain the wurtzite structure and to avoid inclusions of other phases. Cracks must be avoided at all costs since they would inhibit both the in-plane propagation of acoustic waves and in-plane electrical transport.   

To date, \ScAlN layers have mostly been deposited on Si(001) by sputtering.\cite{Campos2018,Qamar2020,Zhu2020,Sundarapandian2023,Lu2018} These layers are uniaxially textured along the [0001] direction and usually exhibit mixed polarity. Recently, \citet{Park2020} have shown that single crystalline Sc$_{0.12}$Al$_{0.88}$N layers can be grown on Si(111) substrates by plasma-assisted molecular beam epitaxy (PAMBE). Even though both textured and single crystalline \ScAlN layers can be used for SAW applications,\cite{Campos2018,Qamar2020,Park2020} single crystalline layers with single polarity are preferred since they generally offer a higher electromechanical conversion efficiency. The same applies for ferroelectric applications since single-polarity layers ensure consistent polarization reversal and offer improved device yield and uniformity.\cite{DWang2022,Wang2023,Zhang2021} The effects of growth temperature on the material quality of \ScAlN have previously been studied by sputtering,\cite{Campos2018,MAkiyama2009,Kobayashi2023} PAMBE \cite{Dinh2023,Hardy2017,Hardy2020,Dzuba2022} and reactive MBE.\cite{Elias} \ScAlN can be produced by sputtering even at room temperature,\cite{Campos2018,MAkiyama2009} while the lowest temperature employed so far in PAMBE is 360\temp.\cite{Hardy2017} Reducing the growth temperature is instrumental in preventing cracks in \ScAlN layers on Si(111).\cite{Lu2018} 

In this work, we investigate the growth of \ScAlN layers on Si(111) by PAMBE, focusing on the effects of growth temperature on the structural perfection of the layers. Particularly, we are concerned about the phase purity of the layers as well as the occurrence of cracks. By successively lowering the growth temperature with increasing Sc content, we are able to obtain crack-free and phase pure wurtzite \ScAlN layers with an on-axis (off-axis) x-ray rocking curve width less than 2° (3°) up to a Sc content of 0.3.

\ScAlN layers are grown by PAMBE on either an $n$-type or semi-insulating 2-inch Si(111) wafer. Before being loaded into the ultrahigh vacuum environment, the Si wafer is ultrasonically cleaned in acetone and isopropanol to remove surface contaminants, and then rinsed with de-ionized water and finally blown dry with a nitrogen gun. Afterward, the wafer is outgassed for 2 hours at 500\temp in a load-lock chamber attached to the MBE system. After being transferred into the growth chamber, the wafer is then heated up to desorb the native oxide at 1000\temp (thermocouple temperature, \Tg). Afterward, the wafer is cooled to 740\temp or the growth temperature given explicitely below. The $(1 \times 1) \rightleftharpoons (7 \times 7)$ surface reconstruction transition, which is known to take place at a temperature of $\approx 860$°C on Si(111),\cite{Kitamura1991May,Miki1992Jul,Hirabayashi1993Apr} is observed in-situ by reflection high-energy electron diffraction (RHEED) at a thermocouple temperature of 830°C.  

The MBE growth chamber is equipped with high-temperature effusion cells to provide the group III metals (99.9999\,\% pure Al, and 99.999\,\% pure Sc). A Veeco UNI-Bulb radio-frequency plasma source is used for the supply of active nitrogen (N$^*$). N$_2$ gas (99.9999\,\%) is as precursor that is further purified by a getter filter. The N$^*$ flux is calculated from the thickness of a GaN layer grown under Ga-rich conditions and thus with a growth rate limited by the N$^*$ flux. \cite{Dinh2023} Prior to \ScAlN growth, we saturate the clean Si(111)-($7 \times 7$) surface with 10 monolayers of Al at \Tg\,=\,740\temp. This Al pre-deposition inhibits the nitridation of the Si surface, and has also been found to be instrumental in producing N-polar AlN on Si(111)-($7 \times 7$).\cite{Dasgupta2009,Ledyaev} Afterwards, the N shutter is opened, and a 20-nm-thick AlN buffer layer (see Fig.\,S1 of the supplementary material for the thickness determination) is grown at the same temperature with an N$^*$/Al ratio of 3.3. During the growth of this buffer layer, the N$^*$-rich conditions assist in gradually consuming the excess Al from the initial pre-deposition step. 

The growth temperature for the subsequent \ScAlN layers is set to 150--740\temp dependong on their Sc content (the standby temperature of the heater is 100\temp). To vary $x$, the layers are grown with different Sc fluxes while keeping the fluxes of Al and N$^*$ constant,\cite{Dinh2023} resulting in a growth rate of 2.8--4.4\,nm/min. The corresponding N$^*$/III ratio thus varies from 3.3 ($x=0$) to 2 ($x=0.3$). Note that at the low temperatures required in our work, metal-stable conditions would inevitably lead to the accumulation of both Al and Sc, resulting in intermetallic precipitates. To ensure a macroscopic compositional uniformity across the wafers, the samples are rotated during growth with a speed of three revolutions per minute. The thickness of the layers of 340\,nm is controlled by in-situ laser reflectometry (see Fig.\,S2 of the supplementary material). The structural properties of the layers are characterized using a high-resolution x-ray diffraction (HRXRD) system (Philips Panalytical X'Pert PRO MRD) equipped with a hybrid two-bounce asymmetric Ge(220) monochromator for the CuK$_{\upalpha1}$ source ($\uplambda$\,=\,\SI{1.540598}\angstrom). $2\uptheta$-$\upomega$ scans are performed with an open detector and without using any receiving slit to maximize the sensitivity to reflections from secondary phases. Azimuthal $\upphi$ scans are used to examine the epitaxial in-plane relationship between the grown layers and the Si substrate. X-ray rocking curves (XRCs) across the on-axis 000\=2 and off-axis 10\=1\=2 reflections are measured with an open detector without using any receiving slit. The lattice parameters of the layers are calculated from the angular positions of the symmetric 000\=2, 000\=4 and asymmetric 10\=1\=5 reflections in 2$\uptheta$ scans, performed using a different HRXRD system with similar configuration but equipped with a PIXcel detector. The surface morphology of the layers is imaged by atomic force microscopy in contact mode (Dimension Edge, Bruker).

\begin{figure}
	\centering
	\includegraphics[width=\columnwidth]{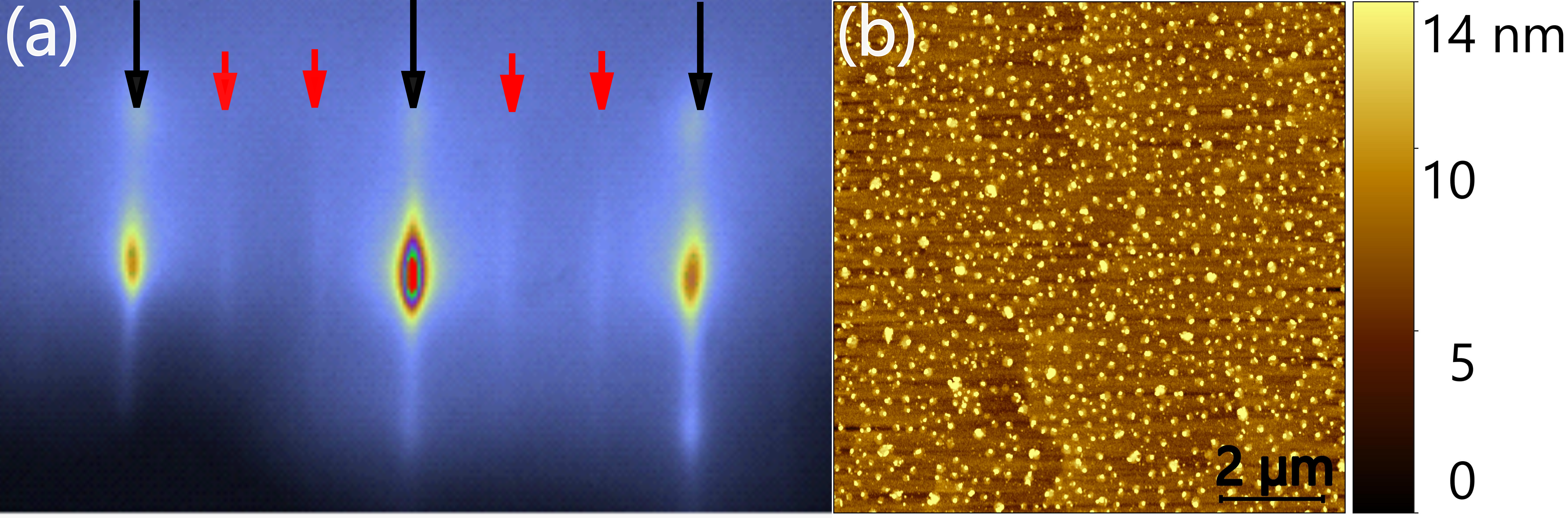}
	\caption{(a) RHEED pattern of the 20-nm-thick AlN buffer layer recorded along the [11\=20]$_\text{AlN}$ azimuth at 100\temp. Long and short arrows point at the integer (bulk) and fractional-order reflections, respectively. (b) $10 \times 10$\,$\upmu$m$^2$ atomic force topograph of this layer with an estimated root-mean-square roughness value of 3\,nm.}
	\label{fig:RHEED}
\end{figure}

To investigate the polarity of the AlN buffer layer (\Tg\,=\,740\temp), RHEED measurements have been performed on the AlN buffer layer at 100\temp. As shown in Fig.~\ref{fig:RHEED}(a), a ($3 \times 3$) pattern is observed, consistent with N-polar AlN.\cite{Smith1998Apr,Dasgupta2009} This assessment is confirmed by KOH etching as shown in Fig.\,S3 of the supplementary material. Hence, it is reasonable to infer that the subsequently grown \ScAlN layers also have N-polarity. The morphology of this AlN buffer layer is characterized by three-dimensional islands on top of terraces separated by atomic steps [Fig.~\ref{fig:RHEED}(b)]. The islands are attributed to the N*-rich conditions and the comparatively low growth temperature.

XRD measurements indicate that the AlN buffer layer is single crystalline with (000\=1) surface orientation. The full-widths at half maximum (FWHMs) of the 000\=2 and 10\=1\=2 XRCs are 0.5° and 0.88°, respectively. Increasing the thickness of the AlN layer from 20\,nm to 340\,nm reduces the FWHMs of the 000\=2 and 10\=1\=2 XRCs to 0.4° and 0.55°, respectively. These values are among the narrowest reported ones for AlN/Si layers grown by PAMBE.\cite{Dasgupta2009,Ledyaev} 

\begin{figure}
\centering
	\includegraphics[width=\columnwidth]{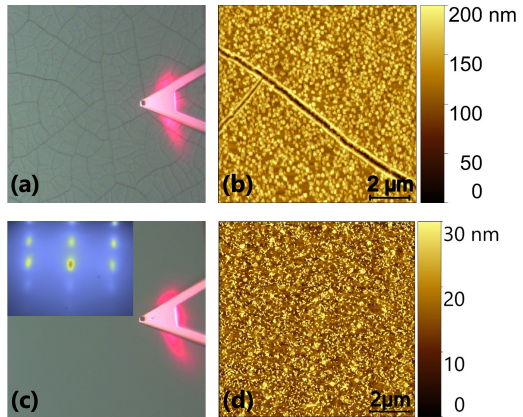}
	\caption{$300 \times 300$\,$\upmu$m$^2$ optical micrographs (taken during AFM measurements while the tip is approaching the surface) and $10 \times 10$\,$\upmu$m$^2$ atomic force topographs of 340-nm-thick \twenty layers grown at 740\temp (top row) and 500\temp (bottom row). Cracks are clearly visible for the former, whereas they are not observed in the latter. The inset of (c) shows a RHEED pattern of the layer recorded along the [11\=20]$_\text{(Sc,Al)N}$ azimuth at 100\temp.}
	\label{fig:AFM}
\end{figure}

\ScAlN layers are grown on this AlN buffer layer with the Sc content $x$ ranging from 0.1 to 0.35. \ten layers grown at the same temperature as the buffer (740\temp) do not show any signs of cracks or phase separation. Cracks are found to first appear in layers with $x \geq 0.2$ grown at this temperature, as shown in Figs.~\ref{fig:AFM}(a) and \ref{fig:AFM}(b). We attribute the crack formation to the tensile stress during cooling from growth to room temperature caused by the large difference in thermal expansion coefficients of \ScAlN and Si. Note that this difference increases with increasing $x$.\cite{Lu2018} 

One obvious measure to avoid cracks caused by thermal stress is to reduce the growth temperature. In fact, as shown in Figs.~\ref{fig:AFM}(c) and \ref{fig:AFM}(d), \twenty layers grown at 500\temp are free of cracks.  The downside of the low \Tg and the N*-rich conditions is the reduced adatom diffusivity and the corresponding increase in surface roughness, as reflected by the transmission pattern observed by RHEED [inset of Fig.~\ref{fig:AFM}(c)] and the AFM topograph shown in Fig.~\ref{fig:AFM}(d). The root-mean-square roughness of this topograph amounts to 7.8\,nm.

\begin{figure}[t]
\centering
\includegraphics[width=\columnwidth]{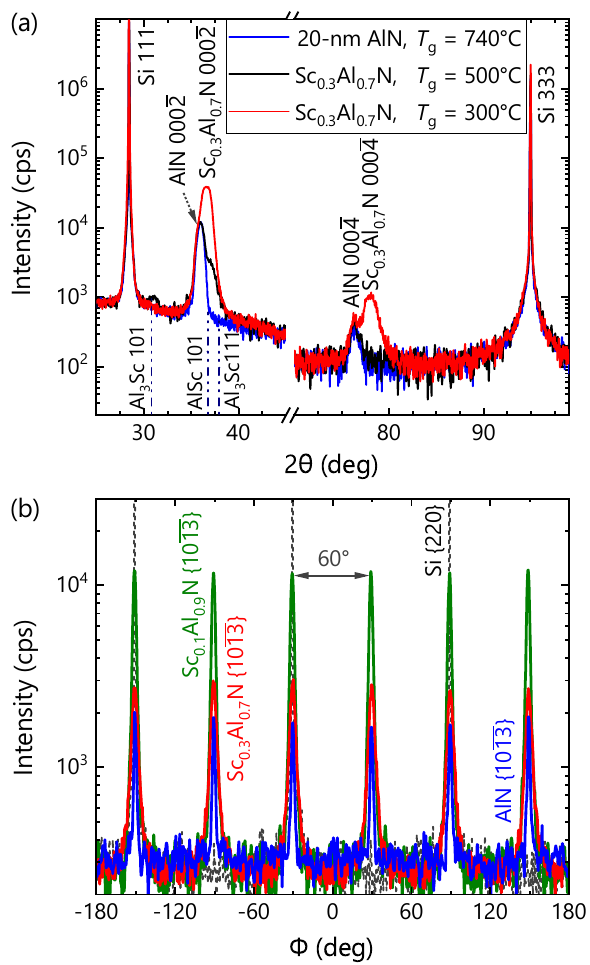}
	\caption{(a) Symmetric \tw XRD scans of the 340-nm-thick \thirty layers grown at 300\temp and 500\temp on Si(111) substrates with 20-nm-thick AlN buffer layers. Note the break of the 2$\uptheta$-axis between 45° to 70°, a range which contains only the Si\,222 reflection at 58.9°. (b) Azimuthal $\upphi$ scans of the Si\,220 reflection and the 10\=1\=3 reflections of AlN, \ten and \thirty.}
	\label{fig:XRD}
\end{figure}  

Phase separation starts to become an issue for $x \geq 0.25$. Specifically, for \twentyfive layers grown at 740\temp, the structural integrity deteriorates due to the formation of intermetallic phases such as AlSc and Al$_3$ScN.\cite{Dinh2023,Elliott} We found that the strategy employed for avoiding cracks, namely, reducing \Tg, helps also to prevent the formation of these secondary phases. In particular, for a Sc content of $x = 0.25$, growth at 500\temp results in layers free of secondary phases. 

However, it turns out that this measure is not sufficient for a Sc content $x \geq 0.30$. Figure~\ref{fig:XRD}(a) shows \tw XRD scans of the 340-nm-thick \thirty layers grown at 500\temp and 300\temp, as well as the scan of the 20-nm-thick AlN buffer layer. The XRD scan of the former sample only exhibits the 000\=2 and 000\=4 reflections of the AlN buffer layer, but not those of the (much thicker) \thirty layer. In addition, the scan exhibits additional reflections due to the presence of both AlSc and Al$_3$Sc inclusions. In striking contrast, the latter sample grown at a furthermore reduced \Tg of 300\temp exhibits intense 000\=2 and 000\=4 reflections, and no reflections indicative of secondary phases. 

\begin{figure}[t]
\centering
\includegraphics[width=\columnwidth]{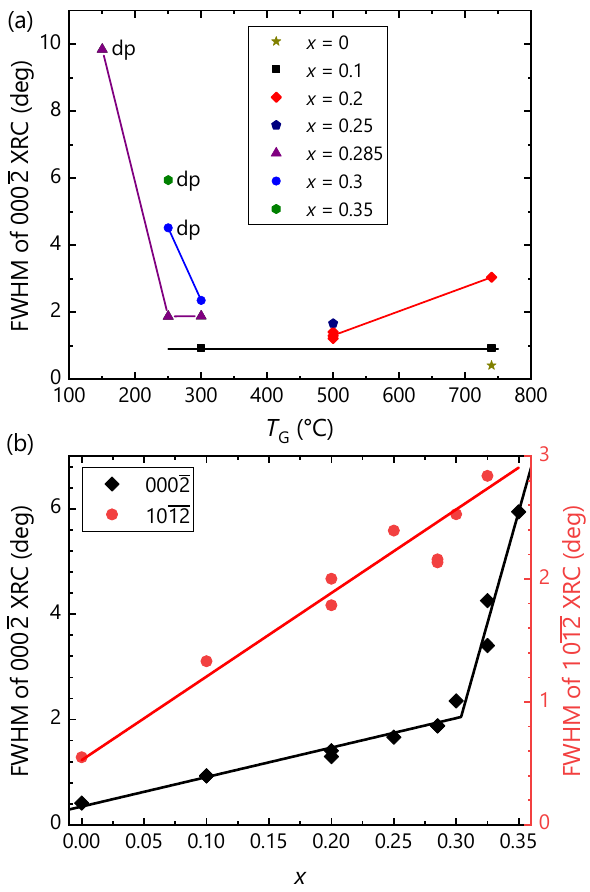}
	\caption{(a) FWHMs of the 000\=2 XRCs for crack-free and phase-pure wurtzite \ScAlN layers grown at different \Tg with a thickness of 240\,nm. For data points designated ‘dp’, the XRCs are dominated by the reflection of the AlN buffer layer. (b) FWHMs of the 000\=2 and 10\=1\=2 XRCs for 340-nm-thick \ScAlN layers grown at optimum \Tg for each $x$. The lines are guides to the eye.}
	\label{fig:XRC}
\end{figure}

To confirm that the \ScAlN layers free of intermetallic phases retain the wurtzite structure, and do not contain additional rocksalt inclusions, we perform azimuthal $\upphi$ scans in skew (quasi-symmetric) geometry of the Si\,220 ($\upomega$\,=\,23.7°, 2$\uptheta$\,=\,47.3°, $\uppsi$\,=\,35.3°, $\upphi$\,=\,0°), AlN\,10\=1\=3 ($\upomega$\,=\,33.0°, 2$\uptheta$\,=\,66.0°, $\uppsi$\,=\,31.6°, $\upphi$\,=\,0°) and \ten/\thirty 10\=1\=3 reflections ($\upomega$\,=\,32.8/32.4°, 2$\uptheta$\,=\,65.6/64.8°, $\uppsi$\,=\,31.1/30.0°, $\upphi$\,=\,0°). As depicted in Fig.~\ref{fig:XRD}(b), the $\upphi$ scan of the \ten and \thirty layers exhibit six maxima separated by 60° with respect to each other, reflecting the sixfold symmetry of the hexagonal wurtzite structure, in contrast to the threefold symmetry of the cubic diamond-structure of the Si(111) substrate. All these layers, including those fabricated at 300\temp, are epitaxial with an orientation-relationship between \ScAlN and Si of [1\=100]$_\text{\ScAlN}$\,||\,[11\=2]$_\text{Si}$ and [11\=20]$_\text{\ScAlN}$\,||\,[1\=10]$_\text{Si}$. No additional reflections pertinent to secondary phases are observed. However, the 10\=1\=3 reflections displayed in Fig.~\ref{fig:XRD}(b) are seen to weaken and broaden with increasing Sc content. Specifically, the FWHMs of the 10\=1\=3 reflections of the AlN buffer layer, \ten and \thirty layers are measured to be 1.8°, 3° and 5°, respectively, reflecting a gradual degradation of the crystallinity with increasing Sc content.

\begin{figure}[t]
\centering
\includegraphics[width=\columnwidth]{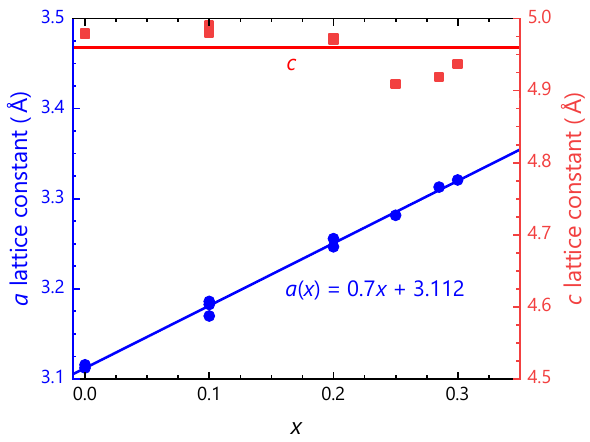}
	\caption{Lattice constants of the \ScAlN layers plotted as function of $x$.}
	\label{fig:lattice}
\end{figure}

For systematically investigating the effect of the Sc content on the layers' out-of-plane and in-plane orientation distribution, we measure on- and off-axis XRCs, respectively, for all \ScAlN layers free of both cracks and phase separation. The results are summarized in Fig.~\ref{fig:XRC}. Note that our assessment of the Sc content may differ from previous ones in the literature.\cite{Dinh2023} 

For a Sc content of  $x=0.1$, the FWHM of the 000\=2 XRC [Fig.~\ref{fig:XRC}(a)] amounts to about 0.9° independent of \Tg, which compares favorably to the 1.2° previously reported for a 400-nm-thick PAMBE-grown Sc$_{0.12}$Al$_{0.88}$N layer on Si(111).\cite{Park2020} As explained above, layers with $x \geq 0.2$ have to be grown at lower \Tg to avoid the formation of cracks and to inhibit phase separation. At the same time, these layers exhibit a significantly narrower out-of-plane orientation distribution as compared to those grown at higher \Tg for the same $x$. In fact, we can keep the FWHM of the 000\=2 XRC lower or equal than 2° up to a Sc content of 0.3 by progressively lowering \Tg to 300\temp.

We note that for $x \geq 0.2$, and \Tg < 300\temp, the symmetric 000\=2 XRCs develop a peculiar shape best described by the superposition of two contributions, a comparatively intense and narrow, and a weak and broad one (see Fig.\,S4 of the supplementary material). The former should not be mistaken for the reflection of the \ScAlN layer. In fact, its FWHM is close to that observed for the bare AlN layer (0.5°), and the narrow (and thus intense) reflection of this buffer layer simply becomes dominant for \ScAlN layers with a broad (and thus weak) reflection. For all such cases, we show the FWHM of the broad component in Fig.~\ref{fig:XRC}(a), labeled ‘dp’ for double peak.

Figure~\ref{fig:XRC}(b) shows the FWHMs of the 000\=2 and 10\=1\=2 XRCs for the 340-nm-thick \ScAlN layers grown at optimized temperatures for each $x$. Both values increase monotonically with $x$ up to 0.3 as also observed previously by other groups,\cite{Hardy2017,Park2020,Hardy2020} but stay below or close to 2°. In contrast to the case of the asymmetric XRCs, the FWHM of the symmetric 000\=2 XRC abruptly increases for $x> 0.3$, reaching values twice higher than those observed for the asymmetric 10\=1\=2 XRC. For wurtzite group III-nitride layers, for which edge ($a$-type) and mixed ($a+c$-type) threading dislocations are typically dominant, this finding is very unusual, and thus points toward a different microstructure. Presumably, cubic precipitations are responsible for tilt boundaries in the \ScAlN matrix, resulting in this large out-of-plane orientation distribution.

Finally, we discuss the in-plane $a$ and out-of-plane $c$ lattice constants of the \ScAlN layers on Si under investigation. Note that for the layers with $x > 0.3$, their lattice constants cannot be measured due to the dominant signals of the AlN layer underneath. The $a$ and $c$ lattice constants of the 20-nm-thick AlN buffer layer and 340-nm-thick AlN layer are comparable, with values of $(3.114 \pm 0.002)$\,Å and $(4.983 \pm 0.004)$\,Å, respectively. These values are is in good agreement with the reported values of relaxed AlN,\cite{Paszkowicz} indicating small residual strains with $\epsilon_\text{xx} \approx 6.5 \times 10^{-4}$ and $\epsilon_\text{zz} \approx 4.2 \times 10^{-4}$. Upon the incorporation of Sc, the $a$ lattice constant is found to linearly increase, while the $c$ lattice constant is almost constant, as shown in Fig.~\ref{fig:lattice}. These findings have been further confirmed by reciprocal space maps of the \ten and \thirty layers, as shown in Fig.\,S5 of the supplementary material. The anomalous changes in the lattice constants of the \ScAlN layers have been first noticed for Sc$_y$Ga$_{1-y}$N and were attributed to the distortion of the wurtzite structure due to the existence of a metastable layered hexagonal phase of ScN.\cite{Farrer,Constantin2004,Constantin2005} More recently, the same behavior has been observed also for \ScAlN.\cite{Deng,Dinh2023}

The $a$ lattice constant of the \ScAlN layers follows a linear relationship $a(x) = (0.7x +3.112)$\,\AA\ as a function of $x$. The slope is slightly smaller than the one measured in our previous work (0.85) for 100-nm-thick Sc$_{x}$Al$_{1-x}$N ($x \leq 0.15$) layers on GaN.\cite{Dinh2023} This result is easily understood: the present layers, given their very large lattice mismatch with Si and their comparatively large thickness, are likely to be essentially fully relaxed, whereas the previous, thinner layers grown on GaN with its significantly smaller mismatch may still be partially strained. Since this strain is tensile in nature for $x \leq 0.1$ and compressive for $x \geq 0.1$, the increase in $a(x)$ is reduced as compared to fully relaxed layers. However, our present data confirm that \ScAlN is lattice-matched to GaN with $x \approx 0.1$\cite{Deng,Dinh2023,vanDeurzen2023Dec} (for an overview of the lattice constants of \ScAlN previously reported in the literature, see Ref.~\citenum{Dinh2023}).

To summarize and conclude, we have investigated the effects of growth temperature on the phase purity and structural perfection of N-polar \ScAlN layers ($0 \leq x \leq 0.35$) epitaxially grown on Si(111) by PAMBE. For obtaining crack-free, epitaxial, and pure wurtzite layers, comparatively low growth temperatures between 500 (for $x \leq 0.25$) and 300\temp (for $0.25 \leq x \leq 0.3$) are required. Still lower growth temperatures result in a severe deterioration of the crystallinity, whereas higher temperatures lead to the formation of cracks and the emergence of intermetallic phases. The realization of phase-pure and crack-free \ScAlN layers on Si  opens up opportunities for integrating the piezo-and ferroelectric properties of \ScAlN with Si, thus adding new functionalities to the Si platform.

\small{See supplementary material for: (1) Symmetric $2\uptheta-\upomega$ x-ray diffraction scan and thickness simulation of a 20-nm-thick AlN(000\=2) layer grown Si substrate; (2) in-situ laser reflectance measurement of a 340-nm-thick Sc$_{0.285}$Al$_{0.715}$N layer grown on 20-nm-thick AlN/Si buffer layer; (3) atomic force topographs of an AlN/Si before and after KOH etching; (4) the 000\=2 and 10\=1\=2 XRCs of the Sc$_{0.285}$Al$_{0.715}$N layers grown at different temperatures; (5) reciprocal space maps of the \ten and \thirty layers taken around the Si\,313 and (Sc,Al)N/AlN\,10\=1\=4 reflections.

We thank Carsten Stemmler for expert technical assistance with the MBE system and Klaus Biermann for a critical reading of the manuscript.}

\bibliography{myBIBss}

\end{document}


\title{Crack-free Sc$_{\boldsymbol{\mathsf{x}}}$Al$_{\boldsymbol{\mathsf{1-x}}}$N(000\=1) layers grown on Si(111) by plasma-assisted molecular beam epitaxy: supplementary material}

\author{Duc V. Dinh}
\email[Electronic email: ]{duc.vandinh@pdi-berlin.de}
\affiliation{Paul-Drude-Institut für Festkörperelektronik, Leibniz-Institut im Forschungsverbund Berlin e.V., Hausvogteiplatz 5--7, 10117 Berlin, Germany.}

\author{Zhuohui Chen}
\thanks{$\dagger$ Deceased on November 4, 2024}
\affiliation{Huawei Technologies Canada Co., Ltd., 303 Terry Fox Drive, Kanata, Ontario, K2K 3J1, Canada.}
	
\author{Oliver Brandt}
\affiliation{Paul-Drude-Institut für Festkörperelektronik, Leibniz-Institut im Forschungsverbund Berlin e.V., Hausvogteiplatz 5--7, 10117 Berlin, Germany.}

\maketitle

\begin{figure}[t]
	\centering
	\includegraphics[width=0.65\columnwidth]{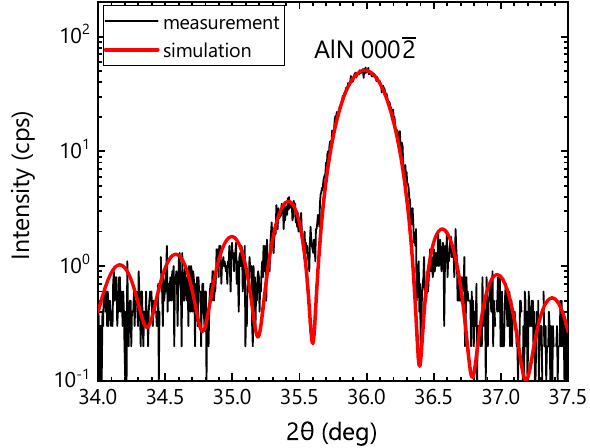}%
	\caption{Experimental and simulated symmetric $2\uptheta-\upomega$ x-ray diffraction scan of a 20-nm-thick AlN(000\=2) layer on Si(111). Thickness fringes around the AlN\,000\=2 reflection indicate the sharp interface between AlN and Si and allow an accurate determination of the thickness of the layer.}
	\label{fig:XRD}
\end{figure}

\begin{figure}[t]
	\centering
	\includegraphics[width=0.65\columnwidth]{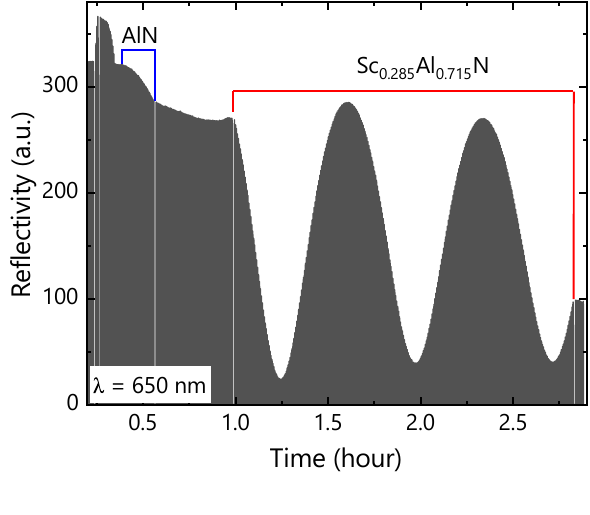}%
	\caption{In-situ laser reflectance measurement (excitation wavelength $\lambdaup$\,=\,650\,nm) of a 340-nm-thick Sc$_{0.285}$Al$_{0.715}$N layer grown on a 20-nm-thick AlN buffer layer on Si(111).}
	\label{fig:LR}
\end{figure}

\begin{figure}[t]
	\centering
	\includegraphics[width=\columnwidth]{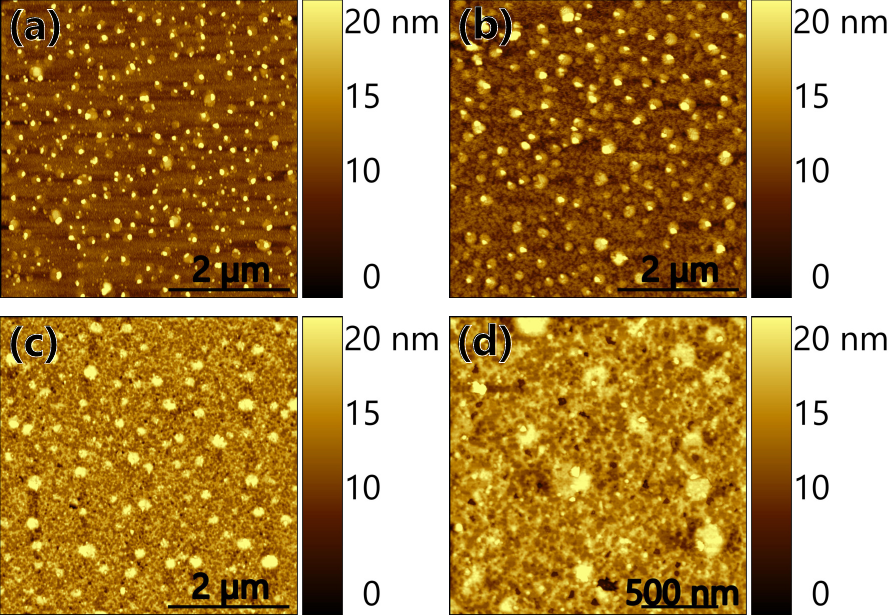}%
	\caption{$5 \times 5$\,$\muup$m$^2$ atomic force topographs of (a) an as-grown AlN sample, (b) after 5-minute KOH-etching, and (c) after 10-minute KOH etching. (d) $2 \times 2$\,$\muup$m$^2$ topograph after 10-minute KOH etching. Surface damage with holes (dark spots) can be clearly seen in (d) and (c), confirming that the layer has N-polarity.}
	\label{fig:KOH}
\end{figure}

\begin{figure}[t]
\centering
\includegraphics[width=\columnwidth]{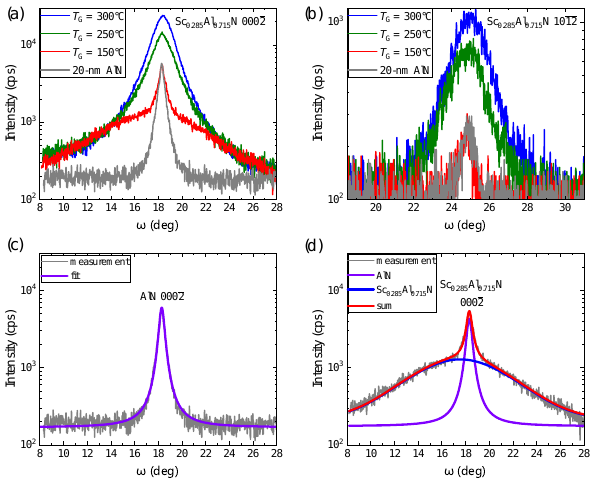}
	\caption{(a) The 000\=2 and (b) 10\=1\=2 x-ray rocking curves (XRC) of 340-nm-thick Sc$_{0.285}$Al$_{0.715}$N layers grown at different temperatures (\Tg). The curves of a 20-nm-thick AlN buffer layer are shown for comparison. (c) Fit of the 000\=2 XRC of the AlN buffer layer using a pseudo-Voigt function. This fit gives a full-width at half maximum (FWHM) of 0.5°, comparable with a value obtained using the commercial X'pert Epitaxy software. (d) A multiple-peak fit of the 000\=2 XRC of the Sc$_{0.285}$Al$_{0.715}$N layer grown at 150\,°C using the pseudo-Voigt function. In this fit, the FWHM of the sharp peak related to the AlN layer underneath is fixed at 0.5°.}
	\label{fig:XRC}
\end{figure}

\begin{figure}[t]
\centering
\includegraphics[width=\columnwidth]{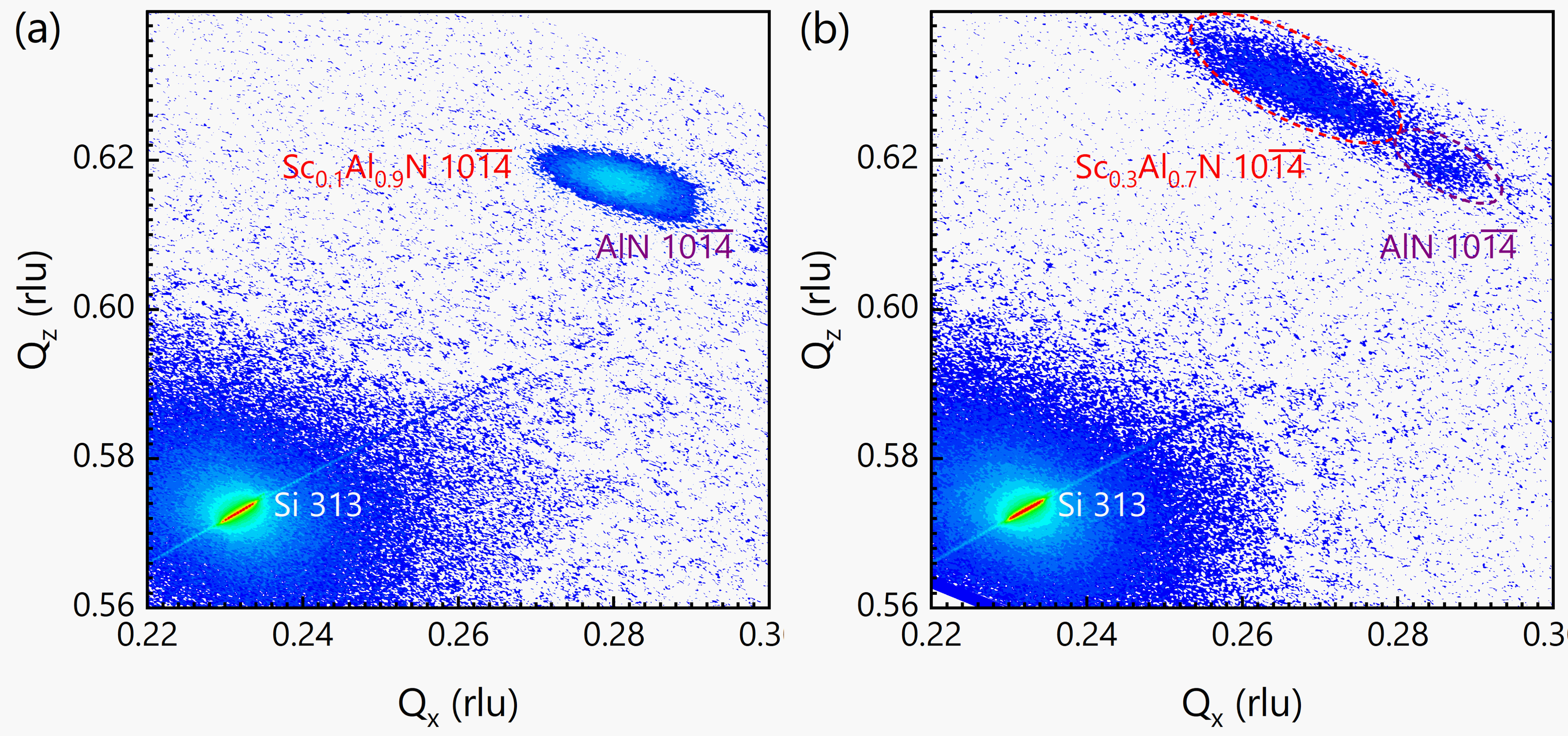}
	\caption{Reciprocal space maps (RSM) of the (a) \ten and (b) \thirty layers grown on Si(111) collected around the Si\,313 and (Sc,Al)N/AlN\,10\=1\=4 reflections. The maps are displayed in reciprocal lattice units (rlu). These maps were recorded with a 1/4° receiving slit. Note that in (a), the AlN\,10\=1\=4 reflection is embedded within the \ten 10\=1\=4 reflection due to the high intensity of the \ten reflection.}
	\label{fig:RSM}
\end{figure}